# High Energy Activity of the Super-Massive Black Hole at the Galactic Center

**Andrea Goldwurm**

Laboratoire Astroparticule et Cosmologie (APC) – Paris
Service d'Astrophysique / CEA – Saclay, 91191 – Gif sur Yvette, France
andrea.goldwurm@cea.fr

## Abstract

The centre of our galaxy hosts the nearest super-massive black hole to the solar system, identified to the compact radio source Sgr A$^*$. High energy experiments have tried in the past to detect the X/gamma-ray emission expected from the accretion of the surrounding material into this super-massive black hole. Only recently however, thanks to the new generation of telescopes, it has been possible to reveal high energy radiation associated with Sgr A$^*$ or its close environment. I will review and discuss in particular the results on the Sgr A$^*$ X-ray flares discovered with Chandra and XMM-Newton, on the central soft gamma-ray source detected with INTEGRAL and on the galactic centre TeV emission revealed by HESS.

### Résumé
Le centre de notre Galaxie abrite le trou noir supermassif le plus proche du système solaire, identifié à la source radio compacte Sgr A$^*$. Plusieurs missions de haute énergie ont essayé, dans le passé, de détecter l'émission X et gamma attendue de l'accrétion de matière environnante dans ce trou noir supermassif. Toutefois seul récemment, et grâce aux nouvelles générations de télescopes, il a été possible de déceler du rayonnement de haute énergie associé à Sgr A$^*$ ou à son environnement proche. Je passerai en revue en particulier les résultats sur les sursauts X de Sgr A$^*$ découverts avec Chandra et XMM-Newton, sur la source centrale de rayons gamma mous détectée avec INTEGRAL et sur l'émission au TeV du centre galactique révélée par HESS.

## 1. Introduction

The galactic nucleus is among the most interesting objects in high-energy astrophysics. Located at about 8 kpc distance, in the direction of the Sagittarius constellation in the southern sky, it hosts the nearest super-massive black hole (SMBH) to the solar system, identified with the compact radio source called Sgr A$^*$. This source is there surrounded by a variety of objects interacting with each other in the dense and complex galactic center (GC) region. Given the relative proximity of this region, the high energy processes generated in this extreme environment, possibly common to other galactic nuclei, can be more easily studied than those in other galaxies.

The GC therefore represents a unique laboratory for modern astronomy and for black hole high energy astrophysics in particular. Black holes (BH) can be studied only trough their effect on the surrounding matter and radiation, and the knowledge of their environment is then of primary importance. Moreover since characteristic timescales of phenomena occurring close to a BH scale linearly with its mass, the timescales involved in SMBH are of the order of minutes and hours instead than milliseconds as in the stellar mass BHs of the galactic X-Ray Binary systems (XRB), and are therefore more easily studied with present day high energy observatories.

The GC SMBH is rather dim at all frequencies, e.g. compared to the powerful BHs in active galactic nuclei (AGN), but such condition appears in fact very interesting for the study of accretion processes. The low level of mass accretion rate implies that radiation is not strongly modified by the surrounding matter (the medium is optically thin) and it is observed at earth as it is produced at the source, with the spectral, polarization and general relativity signatures of the radiation mechanisms and of the BH close environment. Accretion in compact objects generally produces high energy radiation, which is highly penetrating and therefore able to pass through the interstellar medium. Although totally obscured in the visible and ultraviolet by the dust and gas of the galactic plane, the GC can be observed from radio to infrared (IR) and again at high energies (> 1 keV).

For all these reasons, the galactic center has been widely observed in X/gamma-rays and it is still one of the main targets of high energy missions. Some important results have been recently obtained with the new generation of X/gamma-ray observatories: Chandra, XMM-Newton, INTEGRAL and HESS. After an introduction on the GC region (see also the reviews by Mezger et al. 1996, Morris & Serabyn 1996, Goldwurm 2001, Melia & Flacke 2001), I will summarize and discuss the main results concerning the high energy activity of the super-massive black hole of the galactic nucleus.

## 2. The Galactic Centre Region

The spectacular radio image of the GC region, spanning 4º in longitude and 2º in latitude and corresponding to about 600 pc × 300 pc (1º ~ 140 pc at 8 kpc), obtained with the *Very Large Array* (VLA) at 90 cm (La Rosa et al. 2000) shows all the complexity of this peculiar region. The three main extended radio sources, Sgr B, Sgr C and the Sgr A complex in the very centre, include very dense and massive molecular clouds (MC), which, along with the more diffuse matter of the region contain about 10% of the total galactic interstellar molecular gas. Shell-like supernova remnants (SNR) heat the interstellar matter with their non-thermal expanding shells (e.g. G 359.1-005) which appear sometimes interacting with the clouds. While several other non-thermal filaments demonstrate the presence of accelerated particles spiralling around the strong magnetic fields (~1 mG) of the region (like the large filamentary structure known as the Radio Arc), other extended objects (like the so called Bridge) have thermal radio spectra and are in fact HII regions ionized by close hot and young star clusters. The central 50 pc are dominated by the Sgr A radio and molecular complex, formed by few MCs (M-0.02-0.07 and M-0.13-0.06), an expanding SNR, Sgr A East, a central molecular ring and a ionized nebula surrounding the bright compact radio source Sgr A$^*$, the radio manifestation of the central SMBH at the dynamical centre of the Galaxy (Fig. 1).

Sgr A East is a non-thermal radio source composed by a diffuse emission of triangular shape and an inner oval shell (3' × 4', i.e. 7 pc × 9 pc) centered about 50" (~ 2 pc) east of Sgr A$^*$. The shell appears in expansion, compressing the molecular cloud M-0.02-0.07 and probably creating the string of 4 compact HII regions and the OH masers observed around the shell. The first estimates of the shell energy were well above what is expected from the typical release of a supernova (SN), and it was proposed that Sgr A East could be the result of ~ 40 SN or even of the explosive tidal disruption of a star by the SMBH. However X-ray observations indicate now that the energetic content of this object is compatible with a standard SN explosion occurred $10^3$-$10^4$ years ago.

In the inner region a rotating ring of molecular gas and dust, the circum-nuclear disk surrounds Sgr A West, a thermal diffuse nebula with the characteristic shape of a mini-spiral (Fig. 1), also rapidly rotating around the compact source Sgr A$^*$. Sgr A West is ionized by a cluster of hot luminous young and massive stars, centered at about 2" from Sgr A$^*$ and known as the IRS16 cluster. Most of these stars are OB or Wolf-Rayet stars and emit powerful stellar winds which interact with the surrounding medium, form a central cavity and probably feed the SMBH. With individual outflow rates of the order of $10^{-5}$-$10^{-4}$ M$_\odot$ yr$^{-1}$ these stellar winds are probably the origin of the hot gas observed just around Sgr A* in X-rays (Baganoff et al. 2003). Recent results from NIR observations indicate that these very young stars (at radii of about 0.1 pc from the center) seem to be disposed on 2 rotating disks (Paumard et al. 2006). The fact that these stars turn in ordered fashion might have important consequences on the mass and angular moment of the material accreted by the SMBH (Cuadra et al. 2006)

The matter dynamics at radial distance R > 2 pc is dominated by the central core (core radius of ~ 0.4 pc and core density of ~ $10^7$ M$_\odot$ pc$^{-3}$) of the huge star cluster of the galactic bulge, composed by old-middle age stars, whose volume density increases towards the centre with a profile $\propto R^{-2}$. However the large velocities of gas and stars observed in the innermost regions (< 1 pc), especially in NIR by adaptive optics assisted large telescopes like the ESO/NTT, ESO/VLT and the Keck over the last 10-15 yr, must imply the presence of a very compact dark mass, which is now attributed to a super-massive black hole with mass of 3-4 $10^6$ M$_\odot$. In particular studies of proper motions and radial velocities of the *central cluster* stars (a dozen type B stars moving at high velocities within 0.01 pc from Sgr A$^*$) have by now provided precise orbital parameters for 9 of them (Schoedel et al. 2003, Eisenhauer et al. 2005, Ghez et al. 2005). These parameters imply the presence of a dark mass of $3.7 \pm 0.2 \, 10^6$ M$_\odot$ enclosed within a radius < 50 AU. Only a SMBH can explain such densities. The horizon radius of a not rotating BH (the Schwarzschild radius) for such a mass is $R_S$ = 2 GM/c$^2$ ~ $10^{12}$ cm ~ 0.07 AU ~ 15 R$_\odot$ while the Eddington luminosity, the maximum attainable luminosity due to spherical accretion, is of ~ $5 \, 10^{44}$ erg s$^{-1}$.

## 3. Sgr A$^*$: an under-luminous SMBH

The dynamical center of the central star cluster is coincident, within 10 mas, with the bright (~ 1 Jy), compact, variable, synchrotron (flat power law spectrum) radio source Sgr A$^*$. Since its discovery, ~ 30 years ago by Balick and Brown (1974), it has been considered the counterpart of the massive black hole of the Galaxy. The source is linearly polarized at sub-mm frequencies where the spectrum (approximately a power law of energy index ~ -0.3 between 1 and few 100~GHz with both low and high frequency cut-offs) also presents a bump indicating that the emission becomes optically thin above 100 GHz. Sgr A$^*$ proper motion is < 20 km s$^{-1}$ and its intrinsic size, measured at frequencies of 3.5 mm where the interstellar scattering is small, is of the order of 0.1-0.3 mas, about 15-20 $R_S$ (Krichbaum et al. 1998, Shen et al. 2005). The fact that the source is compact and so steady in a region where stars have all very high velocities (500-10000 km s$^{-1}$) indicates that it must be very massive and provides a lower limit of ~ 1000 M$_\odot$ to its mass, excluding the possibility of a stellar object. Thus, as Lynden-Bell and Rees (1971) had predicted 3 years before Sgr A$^*$ discovery, a compact synchrotron radio source at the very centre of our Galaxy reveal the presence of a super-massive black hole in the Galactic Nucleus.

But a black hole surrounded by dense material should accrete and emit high energy radiation as the BHs seen in the XRB and in the AGN. From measures of the stellar wind densities and velocities, it is possible to estimate the capture radius and the accretion rate into the BH and then the accretion luminosity of Sgr A$^*$, it amounts to approximately ~ $10^{43}$ erg s$^{-1}$. If the gas accretes with sufficient angular momentum the flow would set down in an optically thick and geometrically thin accretion disk. Such standard accretion disks have been studied extensively (Shakura and Sunyaev 1973) and it turns out that they have efficiencies of the order of 10%. If present around Sgr A$^*$, such a disk would originate a bright spectral peak in the ultraviolet band similar to the blue bump of AGNs, followed by a steep tail in the soft X-rays. Even if the UVs from the GC are totally obscured, the effect of such a release of energy would be visible and the tails of the emission in NIR and X-rays would be anyway observed. Sgr A$^*$ is instead very dim at all frequencies and the observed bolometric luminosity of this object is well below $10^{37}$ erg s$^{-1}$. This implies an accretion efficiency lower than ~$10^{-6}$ which is extremely low even for the Bondi-Hoyle type of accretion when matter accretes spherically with no transport of angular momentum.

This is the paradox of the low luminosity of the galactic centre massive black hole. In BH binaries and AGN the bulk of the emission is observed at X-rays and therefore important efforts have been done to search for the high energy emission from Sgr A$^*$. This search and the theoretical efforts done to explain the observation results have led to important new developments in accretion theory of compact objects in the regimes of low accretion rate.

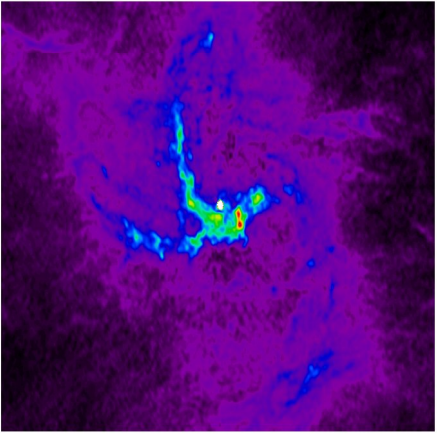 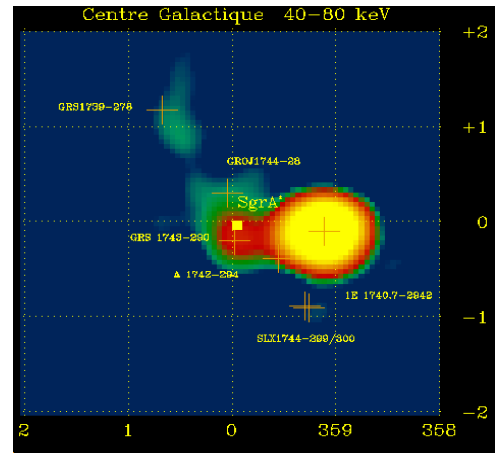

**Fig. 1**  Radio image (~ 1.1' size) of the thermal Sgr A West mini-spiral at 6 cm (courtesy of F. Yusef-Zadeh). The white compact source at the center is Sgr A$^*$.

**Fig. 2**  The SIGMA/GRANAT image of the galactic centre at 40-80 keV, dominated by 1E 1740.7-2942 and a central transient source. Only upper limits were obtained for Sgr A$^*$ (Goldoni et al. 1999).

## 4.     The search for high energy emission from the galactic SMBH

The first detections of high energy emission from the GC direction obtained in the 1970-1980 seemed to support the presence of a SMBH. However as the resolution and sensitivity of the high energy telescopes improved it was realized that the galactic SMBH itself was actually very weak. The first soft X-ray (< 4 keV) images with arcmin resolution were obtained with the Einstein observatory in 1979 (Watson et al. 1981) and showed that the central 20' of the galaxy were dominated by diffuse emission and several weak point-like sources one of which associated to the Sgr A complex. The latter was resolved in 3 weaker sources with Rosat (Predhel & Trumper 1994) and this set an upper limit of only ~ $10^{34}$ erg s$^{-1}$ to the soft X-ray luminosity of the nucleus. On the other hand the BH could still shine in hard X-rays or even at 511 keV, the positron-electron annihilation line, since important fluxes were observed from the general direction of the GC since the early 70s at these energies. After all BH binaries like Cyg X-1 often emit the bulk of their accretion luminosity at > 100 keV. However in the 1990s observations in hard X-rays (3-30 keV) with XRT/SL2 (Skinner et al. 1997) and with ART-P/GRANAT (Pavlinsky et al. 1994) and then in soft gamma rays (30-1000 keV) with SIGMA/GRANAT (Goldwurm et al. 1994, Vargas et al.

1996, Goldoni et al. 1999, Goldwurm 2001) showed that the GC hard X-rays were rather due to the peculiar hard source 1E 1740.7-2942, a BH XRB and microquasar located at > 100 pc away from the centre, and that Sgr A$^*$ was under luminous also at these energies (Fig. 2). The SIGMA telescope also set upper limits on the presence of a point-like source of 511 keV line (Malet et al. 1995) while in those same years OSSE/CGRO showed that the bulk of the 511 keV line emission was not variable but rather diffuse and extended over the whole galactic bulge (Purcell et al. 1997). The pre-Chandra/XMM and pre-INTEGRAL/HESS era ended with the detection of a GC gamma-ray source at energies > 100 MeV with EGRET/CGRO (Mayer-Hasselwander et al. 1998). This source is now positioned slightly aside of the nucleus (~ 0.2°, Hartman et al. 1999) but it could still be linked to Sgr A$^*$.

These high energy results set the Sgr A$^*$ bolometric luminosity to less than $10^{37}$ erg s$^{-1}$ (Goldwurm 2001) and prompted the development of new accretion models where the radiative efficiency is very low. The most popular were those named Advection Dominated Accretion Flows (ADAF) which assume that the energy exchange between protons and electrons is inefficient and electron temperature remain well below the virial temperature reached by the protons. Energy is therefore advected into the BH before being released as radiation (Narayan et al., 1995, 1998). These models have been used for other BH accreting systems with sub-Eddington accretion rates. They predict very low efficiencies, but very hard X-ray spectra extending to the gamma-ray band.

A new era of astronomy was opened with the launch of Chandra and XMM-Newton (1999) X-rays observatories, the launch of the gamma-ray mission INTEGRAL (2002) and the start of operations of the HESS atmospheric Cherenkov ground telescope (2003) for very high energy (VHE) gamma-rays. The Chandra and XMM-Newton surveys of the GC in the 1-10 keV band have shown that the X-ray emission from this region includes few bright, sometimes transient, X-ray binaries probably not associated to the GC (e.g. 1E1740.7-2942 and 1E1743.1-2843); a large population of weak point-like persistent and transient sources (Muno et al. 2003); a diffuse emission with 3 distinct components, a soft thermal one (kT ~ 1 keV), probably SN heated gas, a thermal hot component (kT ~ 8 keV) (Muno et al. 2004) and a non-thermal one characterized by a strong 6.4 keV (Park et al. 2004, Decourchelle et al. 2002). The origin of the hot gas is still a mystery: it cannot be confined by the GC gravitational potential and the heating source is not known (but see Belmont et al. 2005). The nature of the neutral iron fluorescent line at 6.4 keV is also controversial, one possibility being scattering by the MC gas of radiation emitted by an external source in the past, possibly by Sgr A$^*$ itself (e.g. Murakami et al. 2000).

Several SNR, non-thermal filaments and star clusters are also detected. The central 20 pc emission is dominated by the thermal emission from Sgr A East (Maeda et al. 2001, Sakano et al. 2004) which appears as a mixed morphology (non-thermal radio shell with central thermal X-rays) SNR expanding in the dense environment of the GC. Sgr A$^*$ itself appears very weak. Chandra in 1999, with its unprecedented angular resolution of 0.5", further resolved the GC Rosat source and identified what it is now considered the Sgr A$^*$ quiescent emission: a slightly extended source (size ~ 1") with a 2-10 keV luminosity of only 2 $10^{33}$ erg s$^{-1}$ and a steep power-law spectrum (photon index $\alpha$ ~ 2.7) (Baganoff et al. 2003). Such low luminosity and steep spectrum and also the discovery of linear polarization (Aitken et al. 2000) started to challenge the new models. Indeed the polarization at sub-mm frequencies implies an effective accretion rate as low as $10^{-8}$ M$_\odot$ yr$^{-1}$, otherwise the Faraday effect would de-polarize the radiation, making the need of advection less important (Agol 2002).

## 5. Flaring activity of Sgr A$^*$

Certainly the most spectacular Chandra result of the GC observations was the discovery in October 2000 of a powerful X-ray flare from Sgr A$^*$. During this event, of a total duration of 3 hr, the source luminosity increased by a factor of ~ 45 to reach a value of $10^{35}$ erg s$^{-1}$ displaying a hard spectral slope ($\alpha$ ~ 1.3) (Baganoff et al. 2001). XMM-Newton confirmed the presence of such bright hard flares (Goldwurm et al. 2003), and discovered the most powerful one with an increase factor of 180 and, this time, a significantly steeper spectrum ($\alpha$ ~ 2.5) (Porquet et al. 2003, Bélanger et al. 2005). The flare durations (few hours) and the observed short time scales variations (~ 200 s) indicate that the X-ray emission is produced within 20 $R_S$. This cannot be accounted for by the standard ADAF models, for which the bulk of the X-ray emission is produced from the whole accretion flow starting at the accretion radius, and several other models are now considered where non-thermal emission plays a major role.

The model by Liu and collaborators (2002, 2004) assumes that accreting matter circularizes close to the horizon and forms a small, very hot, magnetized Keplerian disk in the inner 10-20 $R_S$ where quasi relativistic electrons produce synchrotron radiation in the sub-mm band and, by inverse Compton, the steep X-ray spectrum. Flares can be produced either by sudden increase in accretion rate or release of magnetic energy. Markoff et al. (2001) instead locate the main energy release at the base of a relativistic compact jet rather than in the accretion disk. Substantial modification of ADAF models (inclusion of outflows, convection and non-thermal component) were also considered and even proposed in relation with the jet models (Yuan et al. 2002, 2003, 2004).

The different models can be adjusted to account for the observed spectral shapes but they predict different correlations between sub-mm, NIR and X-ray fluxes during the events. Since the discovery of the flares a lot of efforts have been directed to obtain detection of flares in different frequency ranges. The first important result of these programs was realized with the detection of NIR flares from Sgr A$^*$ with VLT, Keck and even the HST (Genzel et al. 2003, Ghez et al. 2004, Yusef-Zadeh et al. 2006). NIR flares occur more frequently than X-rays ones (several per day compared to 1 per day) and appear to have steep non-thermal spectra indicating that they are produced by synchrotron emission (Eisenhaeur et al. 2005). Changes in the spectral slope have been observed and a flux-slope correlation (the more intense the harder) seems to be present even during a given event (Gillessen et al. 2006). However the most important diagnostic for modelling would be given by the simultaneous measure of sub-mm/NIR/X spectra of the same flare. Several programs have been carried out to catch flaring activity at different wavelengths. The first simultaneous flare observed in X-rays and NIR was reported by Eckart et al. (2004). However the flare was weak and no clear light curve and spectral shape could be derived from X-ray data. A much brighter one seen simultaneously with XMM-Newton and NICMOS/HST was observed on the 31$^{st}$ August 2004 during a large multi-wavelength campaign on Sgr A$^*$ (Belanger et al. 2005, Yusef-Zadeh et al. 2006). These measures (along with more recent one reported by Eckart et al. 2006) indicate that the NIR and X-ray events are simultaneous and no delay are observed. This implies that the same population of particles must produce the 2 components. Since synchrotron cooling time for producing X-rays in the typical 10 G magnetic field of the accretion flow is much shorter (< 1 mn) than the duration of flares we expect the X-rays to be produced by inverse Compton of sub-mm photons off the transiently accelerated electrons producing the NIR emission via synchrotron. The different spectral slopes of flares could be due to varying electron distribution or magnetic field (Yusef-Zadeh et al. 2006, Liu et al. 2006). Inverse Compton emission could also extend in the gamma-ray regime depending on the above parameters.

Another extremely important issue is the possible detection of quasi periodic modulation during the NIR or the X-ray flares. The first QPOs from Sgr A$^*$ were reported by Genzel et al. (2003) who found in one of the NIR flares a period of 17 mn. If such a period is real and is associated to the Keplerian orbit at the last marginally stable orbit (LSO) of an accretion disk around a black hole it implies that the SMBH is rotating at 50% of the maximum allowed spin. Revisiting X-ray flare variability Aschenbach et al. (2004) found a set of five periods in the power spectra of two X-ray flares. These periods appear in relation with the characteristic frequencies (Keplerian, vertical and radial epicyclical, Lense-Thirring precession) of a disk orbiting a BH with a spin very close to the maximum allowed value. A more detailed timing analysis of the X-ray data however has been recently presented by Bélanger et al. (2006b). These authors found in the Sgr A$^*$ 10 ks X-ray flare of 31$^{st}$ August 2004 a very significant QPO at 22.2 mn (1330 s). Again this period is shorter than the period at the LSO for a Schwarzschild BH and it implies, if associated to stable orbits, a Kerr BH with spin parameter a ~ 0.2 or higher. These authors also re-analyzed all the available XMM data and found not evidence for the modulations reported previously. Although these results are extremely exciting they are still rather controversial and need to be confirmed by other significant measurements. Firm detections of QPO from Sgr A$^*$ with periods in the range 15-30 mn would certainly favor the accretion disk models (vs. jet models) and would provide strong constraints on the mass an spin of the GC SMBH.

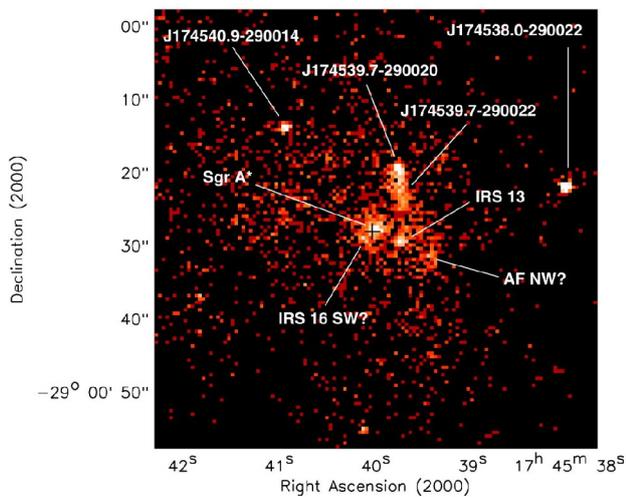
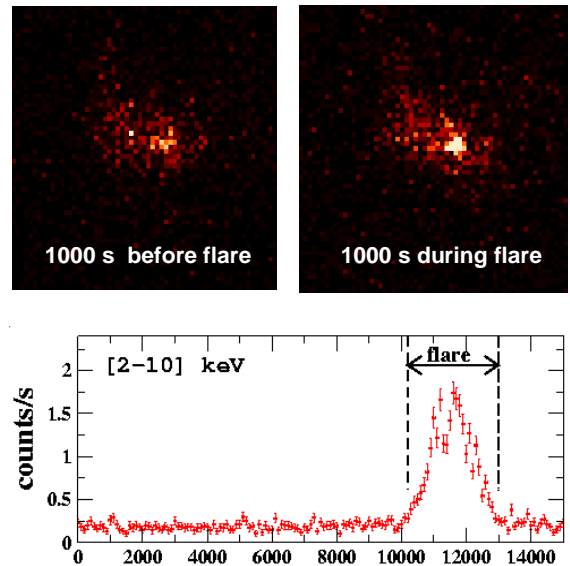

**Fig. 3** Chandra 2-8 keV X-ray image (1.1' size as Fig. 1) of the galactic centre showing Sgr A$^*$ quiescent emission (Baganoff et al. 2003).

**Fig. 4** *Top :* XMM-Newton (MOS) 2-10 keV images (5' size) before (left) and during (right) a Sgr A$^*$ flare (Goldwurm et al. 2003). *Bottom :* XMM-Newton (PN) light curve of the brightest Sgr A$^*$ X-ray flare (Porquet et al. 2003).

## 6. Gamma-Ray emission: INTEGRAL and HESS results

The gamma-ray INTEGRAL observatory monitored the GC region for more than 4.6 Ms effective exposure between 2003 and 2004. The most precise images of the GC ever collected in the 20-600 keV band are those obtained with the IBIS/ISGRI telescope during this survey and were presented by Bélanger et al. (2004, 2006) and Goldwurm et al. (2004) (see Fig. 5). In addition to few bright X-ray binaries INTEGRAL detected, for the first time, a faint and persistent high energy emission coming from the very center of the Galaxy, compatible, within the 1' error radius, with Sgr A$^*$. However the IBIS angular resolution (~ 13' FWHM) does not allow to clearly associate this source (IGR J17456–2901) to the SMBH or to others objects of the dense nuclear region. The lacks of variability and of a bright discrete X-ray counterpart suggest that it is rather a local (size < 10-20 pc) and yet diffuse emission. The INTEGRAL spectrum was compared to the 1−10 keV one obtained from XMM-Newton (partly simultaneous) data, integrating X ray intensities over the region of the IBIS point spread function. The spectra combine well but the thermal plasma of kT ~ 8 keV used to model the bulk of the X-ray diffuse emission cannot explain the data at > 20 keV. The weak transient sources seen by Chandra or XMM in the central arc-minute also cannot explain the observed emission.

A non-thermal component extending up to 120-200 keV with spectral slope of photon index 3 is clearly present and its origin is still unexplained. Simultaneous XMM-INTEGRAL observations performed during the 2004 multi-wavelength campaign dedicated to Sgr A$^*$ are not conclusive on the presence of gamma-ray flares during Sgr A$^*$ X-ray flares since the two observed X-ray events occurred during the intervals when INTEGRAL was in the radiation belts and instruments were off. However, even if the Sgr A$^*$ X-ray flares extend at > 20 keV with their hard slope, they are too sparse to fully account for the gamma-ray source.

In addition, INTEGRAL observed constant hard emission from Sgr B2 (IGR J17475–2822 in Fig.5). This supports the hypothesis that Sgr B2 is a reflection nebula possibly generated by a luminous high energy outburst of Sgr A$^*$ occurred ~ 300 years ago, of which we see today the Compton scattered hard radiation and the fluorescent iron line emission (Revnivtsev et al. 2004).

IGR J17456–2901 could be linked to the very high energy (VHE) gamma-ray emission observed by several Atmospheric Cherenkov Detectors. HESS, the most sensitive and precise of them, reported the presence of a TeV source centered at less than 1' from Sgr A$^*$ (Aharonian et al. 2004) (see Fig. 6). The source is constant and displays a power-law spectrum extending from 300 GeV up to 10 TeV with photon index 2.2. This emission cannot be explained by heavy dark matter particle annihilation and is probably due to interactions of particles accelerated at very high energies. In a following paper (Aharonian et al. 2006) the HESS collaboration also reported TeV diffuse emission closely correlated to the giant molecular clouds of the galactic centre region (Fig. 6). The spectrum and distribution of this emission are consistent with the idea that the central source accelerated in the past the hadrons that are now seen to interact with the molecular gas.

However the mechanism and precise site of the acceleration, the expanding shell of the Sgr A East SNR (Crocker et al. 2005) or the massive black hole itself (Aharonian & Neronov 2005), are not yet identified. The EGRET source observed between 50 MeV and 10 GeV (3EG J1746–2852) located at 0.2º from Sgr A$^*$ seems too far to be the 1~GeV counterpart for the INTEGRAL and the HESS sources, but in this complex region the EGRET data are not conclusive. Gamma-ray emission from the GC has now been clearly detected but its origin and nature are not yet fully understood.

## 7. Perspectives

The Chandra, XMM-Newton, INTEGRAL and HESS monitoring of the GC will continue in the coming months and years, hopefully coupled to NIR, sub-mm and radio correlated observing programs. These programs (and those carried out with the new X-ray observatory Suzaku, launched in 2005) will possibly settle the issue of periodicities in the X-ray flares and will provide measures of the broad band spectra of the Sgr A$^*$ flares and more precise measures of the central high energy emission. In addition the Auger observatory will definitely settle the question of a possible anisotropy of the cosmic rays distribution towards the GC (Crocker et al. 2005) suggested by previous measures but not confirmed by Auger till now (Abraham et al., 2006). Solving the puzzle of the hard X-ray emission observed with INTEGRAL will however necessitate focusing instruments in this energy domain, as Simbol-X expected to fly at the beginning of the next decade (Ferrando et al. 2005). In the near future GLAST will probably unveil the mystery of the EGRET source at the GC and the next generation of ACD detectors (e.g. HESS 2) will map the region at TeV energies with increased precision and will identify which source (Sgr A* or Sgr A East) is at the origin of the high energy hadrons.

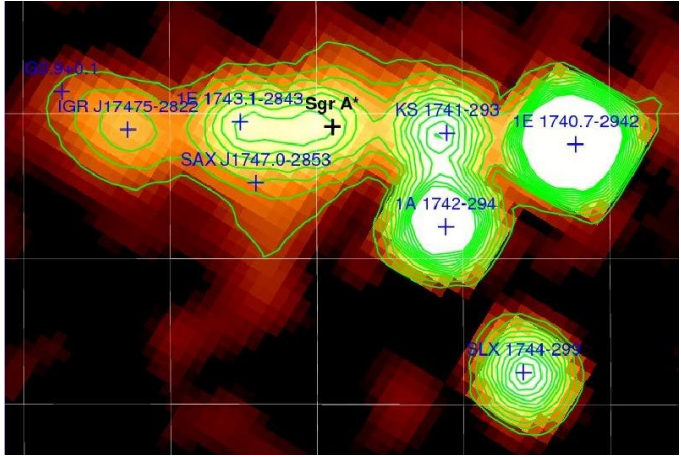

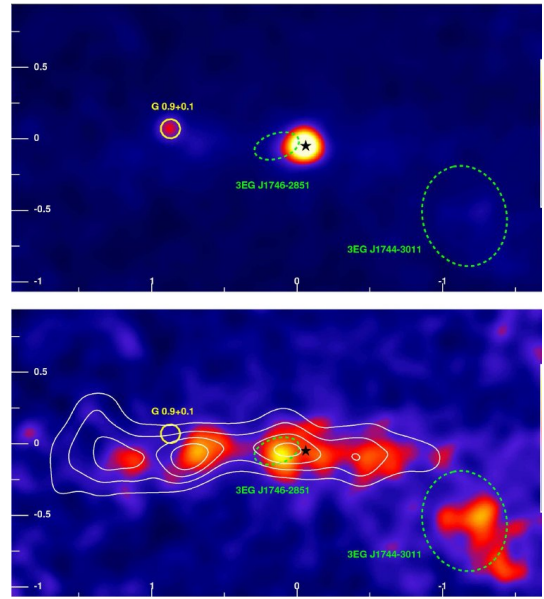

**Fig. 5** INTEGRAL/IBIS 20-30 keV image of the galactic center from the 2003-2004 survey. The image is in galactic coordinates with lines separated by 0.5°. Contours mark iso-significance linearly-spaced levels from 9.5 to 75 σ (Bélanger et al. 2006).

**Fig. 6** HESS 0.4-10 TeV images of the galactic center before (top) and after (bottom) removal of the 2 main point sources (the central source in Sgr A and G 0.9+0.1). Contours are from molecular (CS) radio lines and show the very tight correlation with the diffuse TeV emission (Aharonian et al. 2006).